\documentclass{ametsocV6.1_modified}

\title{Physics-inspired adaptions to low-parameter neural network weather forecasts systems}

\authors{Sebastian Scher\aff{a,b}\correspondingauthor{Sebastian Scher, sscher@know-center.at},
Gabriele Messori\aff{b,c}}

\affiliation{\aff{a}{Know-Center Research GmbH, Graz, Austria}\\
\aff{b}{Department of Meteorology and Bolin Centre for Climate Research, Stockholm University, Stockholm, Sweden}\\
\aff{c}{Department of Earth Sciences and Centre of Natural Hazards and Disaster Science, Uppsala University, Uppsala, Sweden}
}

\abstract{
Recently, there has been a surge of research on data-driven weather forecasting systems, especially applications based on convolutional neural networks (CNNs). These are usually trained on atmospheric data
represented on regular latitude-longitude grids, neglecting the curvature of the Earth. We assess the benefit of replacing the standard convolution operations
with an adapted convolution operation which takes into account the geometry of the underlying data (Spherenet convolution), specifically near the poles. Additionally, we assess the effect of including the information that the two hemispheres of the Earth have ``flipped''
properties - for example cyclones circulating in opposite directions - into the structure of the network. Both approaches are examples
of physics-informed machine learning. The methods are tested on the WeatherBench dataset, at a resolution of $\sim1.4^{\circ}$ which
is higher than many previous studies on CNNs for weather forecasting. For most lead times up to day +10  for 500 hPa geopotential and 850 hPa temperature, we find that using Spherenet convolution or including hemisphere-specific information individually lead to improvement in forecast skill. Combining the two methods typically gives the highest forecast skill. Our version of Spherenet is implemented flexibly and scales well to high resolution datasets, but is still significantly more expensive than a standard convolution operation. Finally, we analyze cases with high forecast error. These occur mainly in winter, and are relatively consistent
across different training realizations of the networks, pointing to flow-dependent atmospheric predictability.}
\begin{document}

%% Necessary!
\maketitle

%%%%%%%%%%%%%%%%%%%%%%%%%%%%%%%%%%%%%%%%%%%%%%%%%%%%%%%%%%%%%%%%%%%%%
% SIGNIFICANCE STATEMENT/CAPSULE SUMMARY
%%%%%%%%%%%%%%%%%%%%%%%%%%%%%%%%%%%%%%%%%%%%%%%%%%%%%%%%%%%%%%%%%%%%%
%
% If you are including an optional significance statement for a journal article or a required capsule summary for BAMS 
% (see www.ametsoc.org/ams/index.cfm/publications/authors/journal-and-bams-authors/formatting-and-manuscript-components for details), 
% please apply the necessary command as shown below:
%
% Significance Statement (all journals except BAMS)
%
%\statement
%	 Enter significance statement here, no more than 120 words. See \url{www.ametsoc.org/index.cfm/ams/publications/author-information/significance-statements/} for details.
%
%% Capsule (BAMS only)
%%
%\capsule
%       Enter BAMS capsule here, no more than 30 words. See \url{www.ametsoc.org/index.cfm/ams/publications/author-information/formatting-and-manuscript-components/#capsule} for details.
%
%% * * If using twocol mode, you will need to use the commands "twocolsig" and "twocolcapsule" in place of "sig" and "capsule"
%%      to ensure that the text box correctly spans across both columns.
%

%%%%%%%%%%%%%%%%%%%%%%%%%%%%%%%%%%%%%%%%%%%%%%%%%%%%%%%%%%%%%%%%%%%%%
% MAIN BODY OF PAPER
%%%%%%%%%%%%%%%%%%%%%%%%%%%%%%%%%%%%%%%%%%%%%%%%%%%%%%%%%%%%%%%%%%%%%
%

\section{Introduction}

Weather forecasting has for decades been dominated by numerical models built on physical principles, the so-called Numerical Weather Prediction Models (NWP). These models have seen a constant increase in skill
over time \citep{bauer_quiet_2015}. Recently, however, there has been a surge of interest in data-driven weather forecasting in the medium-range ($\sim$2-14 days ahead). These have often - but not exclusively, used
neural networks (e.g. \citet{scher_toward_2018,scher_weather_2019-1,dueben_challenges_2018,weyn_can_2019,weyn_improving_2020,faranda2021enhancing,scher_ensemble_2020,rasp2021data,bi2022panguweather, keisler_forecasting_2022, pathak_fourcastnet_2022,chen2023fengwu, lam2023graphcast, benbouallegue2023rise} ), also in combination with physics-based models (e.g. \citet{arcomano2022hybrid}). A historic overview of paradigms in weather prediction, is outlined
in \citet{balaji_climbing_2020}. 
The use of convolutional neural networks (CNNs) (e.g. \citet{scher_toward_2018,scher_weather_2019-1,weyn_can_2019,rasp2021data}) or of a local network that is shared across the domain \citep{dueben_challenges_2018}, dominated in the early data-driven approaches.
What these methods have in common is that they use global data on a regular lat-lon grid. This leads to distortions, especially close to the poles \citep{coors_spherenet_2018}. However, a standard convolution or shared local architecture does not take such distortion into account since it uses a filter whose size is a fixed number of gridpoints (e.g. 3$\times$3). Therefore, the area that the filter sees is not the same close to the Equator and close to the poles. \citet{weyn_improving_2020} have proposed
a solution to this problem via working on a different grid. Specifically, they regrid the data to a ``cubed sphere'' consisting of 6 different
regions. Then, they use a standard convolution operation on each side of the cubed sphere. Additionally, they do not share the weights of the filters globally (as in the original architecture proposed by \citet{scher_toward_2018} and adapted to real world data by \citet{weyn_can_2019}),
but instead use an independent convolution operation for the different sides of the cubed sphere. The weights are shared only for the two polar parts of the cubed sphere, but then ``flipped'' from one pole to the other to account for the different direction of rotation. 

There are also other possibilities for including the spherical nature of the Earth in NN-based weather prediction models. Most of the currently best-performing models use either transformer-based methods \citep{bi2022panguweather} or graph neural networks (GNNs) \citep{keisler_forecasting_2022, lam2023graphcast}.
The approaches with GNNs consider the spherical nature of the earth via mapping into an abstract feature space on an icosahedral grid \citep{keisler_forecasting_2022}, or on grids on a multi-mesh representation \citep{lam2023graphcast}.
The transformer architecture of \citet{bi2022panguweather} uses a "3D Earth-specific transformer". In practice, the spherical nature is dealt with an Earth-specific positional bias in the transformer.

 A drop-in replacement for convolution layers is presented in \citet{esteves2023scaling}. They use a highly optimized form of spherical convolution based on spherical harmonics, and showed it can successfully be integrated into ML models for weather forecasting, having a skill comparable to state-of-the-art architectures.

In this paper, we present an alternative approach to incorporate the spherical nature of the Earth into CNNs. We use the SphereNet architecture, which has previously been proposed for classification
tasks on 360$^{\circ}$-images \citep{coors_spherenet_2018}. Additionally, we test two different approaches for including information on the characteristics of the two hemispheres into our networks. All these approaches can be seen as variants of ``informed machine learning'' \citep{von_rueden_informed_2020, kashinath2021physics},
in which prior knowledge is included into the machine-learning pipeline. In our case, the prior knowledge is that the Earth is spherical (in contrast to the regular data that we provide), and that the dynamics
of the two hemispheres are -- to some extent -- ``flipped'' relative to each other. This information is directly encoded into the structure of the neural network.

Many of the latest transformer and GNN-based weather predictions models come with high technical and computational demands. While these architectures can outperform CNNs, we see a benefit in testing the limits of CNNs with a limited number of parameters, especially for research groups who may not have enough computational capacity to run very large transformer or GNN models. Therefore, the aim of this paper is not to create the best data-driven weather forecasts (and indeed we note that there are published architectures that perform more skillful predictions). Instead, we present two possible improvements to older CNN-based models, and disentangle their individual contributions to forecast skill. The improvements presented here are conceptually simple, and stem from physical reasoning. We believe that the experiments presented here will support future development of physics-oriented ML weather prediction models, despite the fact that large GNN and transformer based models currently have superior skill. 
We use a testbed reanalysis data from the WeatherBench dataset \citep{rasp_weatherbench_2020}, at a resolution of up to 1.4$^{\circ}$. This is higher than many previous studies which used CNNs. Finally, we include an analysis of the events with highest forecast errors. These are important from an end-user point of view, and in NWP models they have elicited significant attention. The occurrence of unusually bad forecasts (``forecast busts'') in NWP models is connected with specific weather situations \citep{rodwell_characteristics_2013-2,lillo_investigating_2017}, and more generally, different weather situations have different predictability
\citep{ferranti_flow-dependent_2015,matsueda2018estimates}. We analyze whether the poorest forecasts of our data-driven forecasts systems are, as in NWP, associated with specific weather situations.

\section{Methods}

\subsection{Data}
\label{sect:Data}
We use data from WeatherBench \citep{rasp_weatherbench_2020}. This is a dataset specifically designed for benchmarking machine-learning based weather forecasts. The subset we use consists
of ERA5 reanalysis data, regridded to a regular lat-lon grid with
two different resolutions: 2.8125$^{\circ}$ (hereafter called ``low-resolution'' or ``lres'') and 1.40625$^{\circ}$ (hereafter called ``high-resolution'' or ``hres''). The following input variables are used: temperature at
850 hPa, geopotential at 300, 500, 700 and 1000 hPa, as well as top-of-the-atmosphere incident solar radiation. As evaluation variables,
we use geopotential at 500hPa (``z500'') and temperature at
850hPa (``t850''). As additional time-invariant input variables, the land-sea mask and orography from ERA5 are used. We use the period 1979--2016 for training and
validation, and 2017--2018 for evaluation (as proposed in WeatherBench). Following WeatherBench,  all results presented in this paper are on the independent evaluation data that is not used for training and validation. The temporal resolution of the data is 6 hours. An overview of the input data is given in table 1.

\begin{table}
\caption{\label{tab:data}Overview of input data variables and spatial and temporal resolutions}
\begin{tabular}{c c c c}
input variable & spatial resolution lres / hres & temporal resolution \\
\hline
temperature 850hPa  & 2.8125$^{\circ}$ / 1.40625$^{\circ}$ & 6h\\
temperature 500hPa  & 2.8125$^{\circ}$ / 1.40625$^{\circ}$ & 6h\\
geopotential 850hPa  & 2.8125$^{\circ}$ / 1.40625$^{\circ}$ & 6h\\
geopotential 500hPa  & 2.8125$^{\circ}$ / 1.40625$^{\circ}$ & 6h\\
top-of-atmosphere incident radiation  & 2.8125$^{\circ}$ / 1.40625$^{\circ}$ & 6h\\
orography &  2.8125$^{\circ}$ / 1.40625$^{\circ}$ & time-invariant\\
land-sea mask &  2.8125$^{\circ}$ / 1.40625$^{\circ}$ & time-invariant\\

\end{tabular}
\end{table}

\subsection{Spherenet Convolution}

In normal convolution, for each gridpoint, a fixed number of gridpoints
in the vicinity are sampled (for example a 3$\times$3 box centered
on the gridpoint). For data on a globe (such as global atmospheric
data) represented on a regular grid, this leads to distortions except
very close to the equator. Indeed, a fixed neighborhood defined via the
number of gridpoints corresponds to rectangles of differing size,
depending on latitude. One way to remedy this is through the use of
adapted convolution filters. \citet{coors_spherenet_2018} proposed the SphereNet architecture in neural networks for image detection in spherical images.

\begin{figure}

\includegraphics[width=1\textwidth]{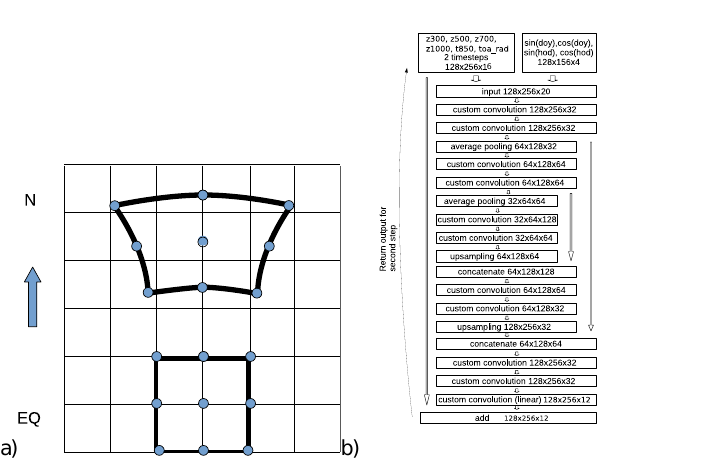}
\caption{\label{fig:sketch-of-the}a) Sketch of the principle of Spherenet
convolution. Shown is a  3$\times$3 kernel (corresponding
to 3$\times$3 gridpoints at the equator). b) Overview
of our neural network structure. Dimensions are for the high-resolution data.}

\end{figure}

With this method, instead of a box of fixed size in gridpoint space,
each gridpoint is assigned a rectangle with fixed size in real space.
Since the positions of available gridpoints and the target coordinates
in this fixed-size box do not necessarily coincide, the target can
also be an interpolation of up to four gridpoints.
The principle is sketched in Fig. \ref{fig:sketch-of-the} a).

In normal convolution operations in neural networks, the kernel is
always made up of exact gridpoints (e.g. a 3$\times$3 box corresponding
to 9 gridpoints). Our approach requires interpolation, and thus our
method also allows the use of a non-integer number of gridpoints in
the kernel. For example, we could use the distances -1, -0.5, 0, 0.5, 1
along the longitude direction, creating a kernel of 5 gridpoints. These 5 points then cover, with equal spacing, the distance that is normally covered by three kernel points (this distance that is is, in km,  the distance of three gridpoints at the equator).

An alternative method for adapting convolution operators to the Earth's approximately spherical geometry that would seem promising in our case was proposed by \citet{boomsma2017spherical}, who presented convolution on a cubed sphere, where the sphere is divided into 6 parts. This is the same approach as in \citet{weyn_improving_2020}, but with weight-sharing across all 6 parts. The main limitation of this approach is that translational invariance of the convolution filters is lost. Specifically, in our application, the filters would not necessarily be aligned along the longitude and latitude circles. This is a potential problem, as there is physical meaning to these (a gradient in a certain variable along the latitude dimension is physically not the same as along the longitude dimension).

\citet{DBLP:journals/corr/abs-1801-10130} compute spherical convolution via FFTs, resulting in full rotational invariance. As noted by \citet{coors_spherenet_2018} this is not always a desired property, as it assumes that all directions (e.g. in our case  along longitude, along latitude or any other direction) are equivariant. As we discussed above, in our setting this is not physically meaningful. 
\citet{eder2019mapped} presents an approach that generalizes the method from  \citet{coors_spherenet_2018} to any type of structured data. \citet{Jiang2019} present an approach that works for unstructured grids.

Due to the limitations of the methods from \citet{boomsma2017spherical} and \citet{DBLP:journals/corr/abs-1801-10130} discussed above,  and the fact that for our setting we don't need the generalized approaches from \citet{eder2019mapped}, and \citet{Jiang2019},  we opted for using the approach from \citet{coors_spherenet_2018}. We will refer to this approach as Spherenet convolution.

\subsubsection{Implementation}
\label{sect:implementation}

Since \citet{coors_spherenet_2018} have not provided details on their
technical implementation, and since their code is not publicly available,
we have designed our own implementation of Spherenet. In this
section we use the word ``tensor'' as it is used in computational
packages such as tensorflow, thus interchangeably with ``array''.
Therefore, not everything referred to as a tensor here is necessarily
a tensor in the strict mathematical sense of the term.

We have implemented Spherenet with the following steps
(the channel dimension of the neural network is omitted here for simplification):
\begin{enumerate}
\item we start with a (fixed) filter kernel $\vec{K}$ of length $n,$ consisting
of $n$ pairs of lat-lon distances $\Delta p_{i}=\left(\Delta y_{i},\Delta x_{i}\right)$,
corresponding to gridpoints at the equator. A 3$\times$3 kernel without
fractional distances for example would be\\ $\left[\left(-1,-1\right),\left(-1,0\right),\left(-1,1\right),\left(0,-1\right),\left(0,0\right),\left(0,1\right),\left(1,-1\right),\left(1,0\right),\left(1,1\right)\right]$.
\item for each of the $N\times M$ input gridpoints $p=\left(x,y\right)$
in the regular grid, we compute $n$ pairs of (potentially non-integer)
coordinates $p'=\left(y'_{i},x_{i}'\right)$, corresponding to the
$n$ points in the kernel $\vec{K}$, transformed for the current
position of $p$ on the globe with the following equations. $x$ and $y$ are in gridpoint coordinates, and $\phi$ and $\theta$ are in radians.

\begin{equation}
\phi'=\arcsin \nu\sin \phi+\frac{\Delta \phi  \sin \nu\cos \phi}{\rho} 
\end{equation}

\begin{equation}
\theta' = \theta +\arctan\left( \frac{\Delta x \sin\nu}{\rho} \cos\phi\cos\nu-\Delta y \sin\phi \sin\nu \right)
\end{equation}
\begin{equation}
    \rho=\sqrt{\left(\Delta\rho\right)^2+\left(\Delta \theta\right)^2}
\end{equation}
\begin{equation}
    \nu=\arctan\rho
\end{equation}
\begin{equation}
    \Delta\phi=\Delta y \frac{\pi}{N}
\end{equation}

\begin{equation}
\Delta\theta=\Delta x \frac{2\pi}{M}
\end{equation}

\begin{equation}
y'=\phi' \frac{M}{\pi}+\frac{N-1}{2}  
\end{equation}
\begin{equation}
x'=\theta' \frac{M}{2\pi}  
\end{equation}

 with latitude $\phi$ of the central point. The transformed coordinates
$\phi'$ and $\theta'$  are in regular lat-lon coordinates, and $x'$ and $y'$ the corresponding coordinate indices on the regular lat-lon grid. The transformed coordinate indices for
each gridpoint and kernel points are combined in a coordinate tensor
$\hat{A}$ of shape $N\times M\times n$. 
\item the input $\vec{x}$ data is flattened to $\vec{x}_{flat}$ with shape
$L=N\cdot M$, and the coordinate tensor $\hat{A}$ is flattened to
a tensor of shape $L\times n$, with the coordinates transformed to
flattened coordinates.
\item a sparse interpolation tensor $\hat{L}$ of size $L\times L$ is created,
and filled with the target coordinates in such a way that multiplying
the flattened input data $\vec{x}_{flat}$ with the interpolation
tensor results in the expanded input data $\vec{x}_{exp}=\hat{L}\vec{x}$
with shape $L\times n$. $\hat{L}$ is implemented as a sparse tensorflow
tensor. This implementation allows the use also on very large grids
(large $L$), as only the non-zero components are kept in memory.
\item on $\vec{x}_{exp}$, a standard 1-d convolution with kernel size $n$
(as implemented in major neural network libraries such as tensorflow)
can now be applied, resulting in $\vec{x}_{out}$ with size $L$,
which is then unflattened to shape $N\times M$
\end{enumerate}
The steps 1 and 2, and the computation of the interpolation tensor, need
to be performed only once (when setting up the network). $\hat{L}$ is
stored in memory for all subsequent operations.

At gridpoints close to the poles, kernel-points can ``pass'' through
the pole. For these points, not only the longitude, but also the latitude
is adjusted. For example, on a 1x1$^\circ$ grid, a kernel point that without
this adjustment would correspond to the impossible point 90.5$^\circ$N 0$^\circ$E
will be set as 89.5$^\circ$N 180$^\circ$E. With this, the ``polar problem'' of regular
grids is eliminated.

\subsection{Neural network architecture}
\label{subsect:NN_architecture}

We use a neural network architecture based on that proposed in \citet{weyn_improving_2020},
namely a U-net architecture. \citet{weyn_improving_2020} however
do not use data on a regular grid, but on a cubed sphere, consisting
of several regular grids. They further use two sets of weights, across six different regions. We use the same architecture,
but with each of their special convolution layers replaced by a standard
convolution, Spherenet convolution and/or hemisphere-wise convolution
(see below). The network structure is shown in Fig. \ref{fig:sketch-of-the}
b). Our networks are implemented in tensorflow  \citep{martin_abadi_tensorflow:_2015} using the dataset api
with tensorflow record files, resulting in an implementation that should also scale to datasets with higher resolution than the ones used here.

In addition to the input variables from ERA5 discussed in Sect. 2.a, 
day of the year (``doy'') and local hour of the day (``hod'', different for each longitude band) are used
as additional input variables. Since these are ``circular'' variables,
each of them is converted to two variables. Using two variables should make it easier for the model to learn the circular relationship (e.g. that doy 1 and doy 365 are adjacent days):

\begin{eqnarray}
doy1 & = & \sin\left(\frac{2\pi}{365}doy\right)\\
doy2 & = & \cos\left(\frac{2\pi}{365}doy\right)\\
hod1 & = & \sin\left(\frac{2\pi}{24}hod\right)\\
hod2 & = & \cos\left(\frac{2\pi}{24}hod\right)
\end{eqnarray}

These 4 scalars are extended to the grid-resolution of the data and
added as additional channels. The input of the networks is comprised
of two timesteps of the 6 ERA5 inputs (resulting in 12 input channels), but the
additional 4 variables (doy1, doy2, hod1, hod2) are provided only once, resulting in 12+4=16
input channels. The outputs of the networks are 2 timesteps of the
input variables, without the additional variables (thus 12 channels).
One forecast step is made of two consecutive passes through the network,
via feeding the output back to the input, resulting in a 24 hour forecast.
For details, see \citet{weyn_improving_2020}.

For consecutive forecasts (longer than 24 hours), hod is not updated,
since each forecast step is 24 hours. We also choose not to updated
doy, since the forecast length of 10 days is very short compared to
seasonal variations. 

\subsubsection{Base architecture}
Our base architecture without Spherenet convolution uses normal convolution with wrapping on the sides.
Along the longitude direction the convolution is ``wrapped'' around,
so there is no artificial boundary. At the poles the grid is wrapped over the pole: the northern neighbour of a point on the northernmost latitude band is the point on the same latitude band but with 180$^\circ$ shifted longitude. The kernel size of the convolutions
is 3$\times$3. Note that for the different resolution datasets, this corresponds to different spatial extents in the input data.

\subsubsection{Spherenet convolution architecture}

The Spherenet convolution architecture is the same as the base architecture,
except that each convolution operation is replaced by a Spherenet
convolution operation. Since the convolution deals both with the poles
and the longitude-wrap, no padding is applied. We use a 3 × 3 kernel,
just as in the base architecture. 

\subsubsection{Hemispheric convolution}

We use two related approaches for incorporating the fact there there
are 2 hemispheres into the networks. In the first, we use separate
(independent) convolution operations (with separate weights) for each
hemisphere. The data is split at the equator. For the architecture
without Spherenet convolution, the first row of the other hemisphere
is added as padding for the boundary of the convolution, and the padding at the pole is handled in the same way as in the base architecture. When using
Spherenet convolution together with hemispheric convolution this is
not necessary, as this is included in the interpolation for the Spherenet
convolution. Then, on each hemisphere, a convolution operation is
performed. These will be referred to as ``hemconv'' and ``sphereconv\_hemconv'', respectively.
In the second approach, the same convolution operation is used for
both hemispheres, with the filter ``flipped'' along the lat dimension
for the second hemisphere. This will be referred to as ``hemconv\_shared''
and ``sphereconv\_hemconv\_shared''. This approach is a variant
of the inclusion of ``invariances'' into the neural network in the
terminology of \citet{von_rueden_informed_2020}. The rationale behind this flipping approach is that many properties of the atmospheric circulation have flipped properties on one hemisphere compared to the hemisphere. For example, many variables -- including temperature and geopotential -- have significant pole-to-equator gradients which change sign between the hemispheres. Additionally, mid-latitude weather systems spin in opposite directions in the two hemispheres which manifests itself, amongst others, in the temperature and geopotential fields. Note that when sharing the weights, but not flipping the weights, one would end up with the base architecture again.
For both hemconv and hemconv\_shared (and sphereconv\_hemconv and sphereconv\_hemconv\_shared, respectively), for each convolution operation, the same convolution depth (=number of filters) is used. Therefore, the networks of hemconv and sphereconv\_hemconv have twice the number of parameters compared to the other architectures.

\begin{table}[]
    \centering
    \caption{Overview of network architectures}
    \label{tab:netoverview}
    \begin{tabular}{c|c}
       architecture  & parameters  \\
       \hline
       basenet  & 336,040 \\
       sphereconv & 336,040 \\
       hemconv & 671,816 \\
       hemconv\_halfsize & 336,040 \\
       hemconv\_shared & 336,040 \\
       sphereconv\_hemconv & 671,816  \\
       sphereconv\_hemconv\_shared & 336,040 \\
       basenet\_latlon & 336,040 \\
    \end{tabular}

\end{table}

\subsubsection{Additional experiments}
In the addition to using all architectures described above, we made two additional experiments on the low resolution data. In the first experiment  we added latitude and longitude grids as constant input features to the input data. Latitude was scaled to [0,1], while the longitude $\theta$ (in radians) was, just as day of the year and hour of the day, presented as two variables 

\begin{eqnarray}
lon1 & = & \sin\left(\theta\right)\\
lon2 & = & \cos\left(\theta\right)\\
\end{eqnarray}

This approach we call base\_latlon.
In the second experiment, called hemconv\_halfsize, we use hemispheric convolution without filter sharing, but with half the number of parameters per hemisphere compared to the base architecture. Whit this, the network in total has the same number of parameters as the base-architecture and all other architectures except for hemvonv\_shared and sphereconv\_hemconv\_shared.
An overview of network architectures and the number of neural network parameters is given in table \ref{tab:netoverview}.

\subsubsection{Network training}

For the training, the data from WeatherBench is converted to the tensorflow-record format. Each network is trained 4 times with different random seeds to account
for the randomness in the training. Each training realization is evaluated
separately, and throughout the paper the average of the errors and
skill scores is shown. The same architecture is used  for both high
and low resolution data. Only the input size is adjusted according
to the resolution. Since the architecture is a pure convolution architecture,
the number of parameters (weights) is independent of the input size,
and thus both for high-resolution and low-resolution the same number of parameters is used
(336,040 for architectures with same weights for both hemispheres,
and 671,816 for the architectures with independent weights for each
hemisphere). We train the networks first for 100 epochs, and then over an additional 50 epochs with early
stopping (stopping after no increase in skill at 10\% of the training
data left out for validation). The latter step is done to prevent overfitting of the model. 

\subsection{Forecast evaluation}

We use both Root Mean Square Error (RMSE) and Anomaly Correlation
Coefficient (ACC), which are also the two evaluation measures used in WeatherBench.

RMSE is defined as
\begin{equation}
RMSE=\overline{\left(fc-truth\right)^{2}}
\end{equation}

with the overbar representing latitude-weighted area and time mean, and
ACC as
\begin{equation}
ACC=corr(fc-clim,truth-clim)
\end{equation}

with the correlation computed with latitude weights and $clim$ the
time-mean over all forecasts. For further details on the calculations, we refer the readers to
\citet{rasp_weatherbench_2020}.

\section{Results}
\subsection{Evaluation of the different forecast architectures}
\label{sect:evaluation}

\begin{figure}
\includegraphics[width=1\textwidth]{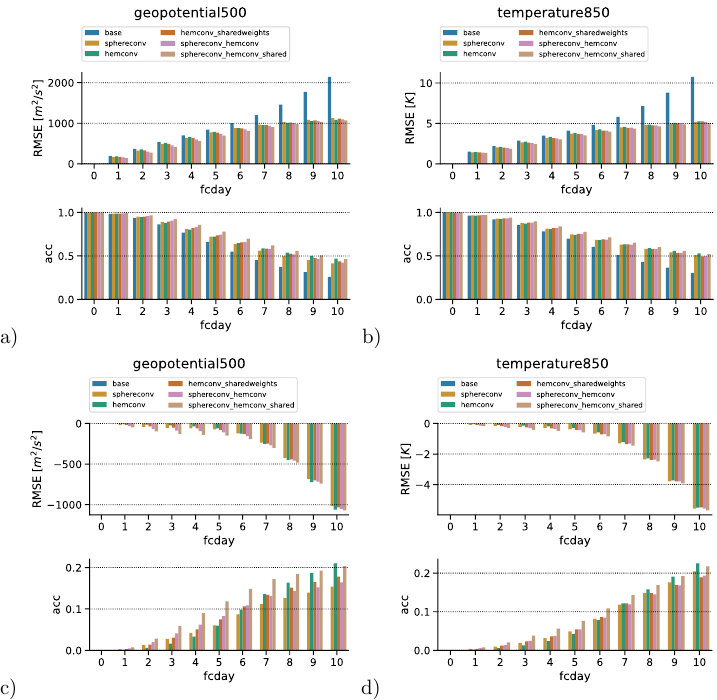}
\caption{\label{fig:forecast-skill-hres}Forecast skill (RMSE and ACC) for
all high-resolution (1.4$^{\circ}$) architectures for geopotential at 500hPa
and temperature at 850hpa. a,b: absolute values; c,d: RMSE and ACC of all architectures minus RMSE and ACC of base architecture. For all plots: The ordering of the bars is the same as the ordering in the legend when read column-wise. The first column top to bottom corresponds to the first four bars, the second column to the remaining four bars.}
\end{figure}

\begin{figure}
\includegraphics[width=1\textwidth]{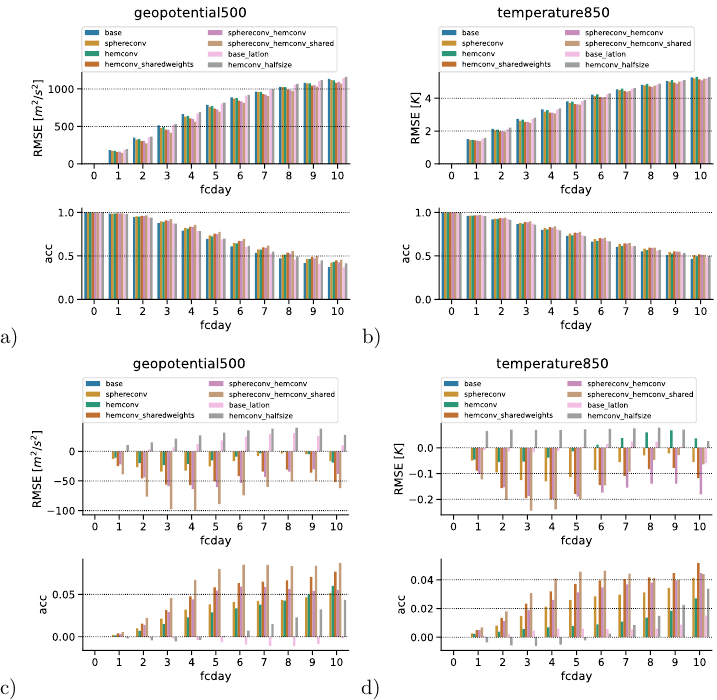}
\caption{\label{fig:forecast-skill-lres} Forecast skill (RMSE and ACC) for
all low-resolution (2.8$^{\circ}$) architectures, for geopotential at 500hPa
and temperature at 850hpa. The figure includes the two additional experiments only performed on the low resolution data:  namely providing latitude and longitude values as constant feature channels ("base\_latlon") and hemispheric convolution with only half the number of parameters per hemisphere ("hemconv\_halfsize") a,b: absolute values; c,d: RMSE and ACC of all architectures minus RMSE and ACC of base architecture. For all plots: The ordering of the bars is the same as the ordering in the legend when read column-wise. The first column top to bottom corresponds to the first four bars, the second column to the remaining four bars.}

\end{figure}

We start by looking at global average RMSE and ACC, shown in Figs.
\ref{fig:forecast-skill-hres} and \ref{fig:forecast-skill-lres} and
Table \ref{tab:baseline-scores}. The upper panels of the figures
show absolute values, whereas the lower panels show the differences relative
to the base architecture. The base architecture has the lowest skill at all lead times and for all resolutions, skill metrics and variables considered here (we do not include here the two additional experiments described in Sect. 2\ref{subsect:NN_architecture}), except for lead times of 6--8 days for RMSE of the low-resolution 850 hPa temperature.
Sphereconv improves over the base architecture for all cases. This
holds for all lead-times, both resolutions and both for RMSE and ACC. In many cases, the improvement is however relatively small. 

The hemconv architecture also improves systematically on the base architecture, with the exception of the above-mentioned lead times of 6--8 days for RMSE of the low-resolution 850 hPa temperature. Depending on the variable, resolution and lead time, it can also perform better than than the sphereconv architecture on the high-resolution
data. Taking as reference RMSE at lead times of 3 or 5 days (Table \ref{tab:baseline-scores}), sphereconv systematically outperforms hemconv. Combining the two (sphereconv\_hemconv), the forecast performance generally improves further over both sphereconv and hemconv taken individually. The hemispheric convolution architecture with shared
(flipped) weights (hemconv\_shared) generally outperforms the hemispheric convolution
architecture with independent weights. The same holds when comparing Spherenet convolution combined with hemisphere-wise convolution with shared weights (sphereconv\_hemconv\_shared) to sphereconv\_hemconv. While no single method is the best for all cases considered, in most cases combining Spherenet convolution with hemisphere-wise convolution and sharing the flipped weights leads to the best results. This is illustrated in Table \ref{tab:baseline-scores}, where the lowest RMSE for each variable, lead time and resolution (not considering IFS) is highlighted in bold.

\begin{figure}
\includegraphics[width=1\textwidth]{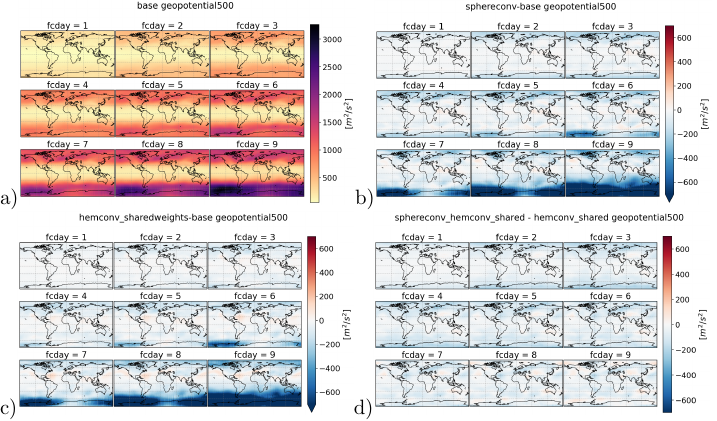}
\caption{\label{fig:a) spatial}a) Spatial distribution of RMSE of geopotential at 500 hpa $[m^{2}/s^{2}]$ for forecasts days 1-9
of the base architecture; b) difference between RMSE of sphereconv
and RMSE base; c) difference between RMSE of hemconv\_shared and RMSE of base; and d) difference
between RMSE of sphereconv\_hemconv\_shared and RMSE of hemconv\_shared}

\end{figure}

We now turn to the spatial distribution of RMSE. We focus on z500 as this provides a more direct link to the atmospheric circulation than temperature (Fig. \ref{fig:a) spatial}).
Panel a) shows the error of the base networks at different lead times
for the high-resolution data, with increasing leadtime from upper left to lower
right. The error pattern follows the typical error patterns of medium
range NWP forecasts, with lowest predictability in the storm-track
regions and in the Southern Hemisphere (e.g. \citet{scher_how_2019}). As expected, the error grows
with increasing lead-time, with no dramatic changes in the spatial patterns. 

More interesting is the difference between the spherconv and the base architecture (Fig. \ref{fig:a) spatial}b) and between hemconv\_sharedweights and the base architecture (Fig. \ref{fig:a) spatial}c). Both clearly improve the forecasts in the high latitudes, and in particular in the Southern Hemisphere, with increasing improvement with increasing lead times. We finally consider the difference between sphereconv\_hemconv\_shared
and hemconv\_shared (Fig. \ref{fig:a) spatial}d). Up to forecast day 5, the Spherenet
convolution clearly improves the forecasts around both poles. For longer lead times, the Spherenet convolution still provides an improvement, but this now appears to be largest in the mid-latitudes. Results are similar for t850 (Fig.
A1), except that the forecast improvements in the mid-latitudes for both sphereconv and hemconv\_shared are larger relative to those in the poles than for z500. Moreover, sphereconv\_hemconv\_shared presents a slightly lower skill in some mid-latitude regions compared to hemconv\_shared. 

Finally, we turn to the two additional experiments carried out on the low resolution data: including latitude and longitude as additional constant channels (base\_latlon), and hemispheric convolution without shared weights, but with half the number of weights per convolution (hemconv\_halfsize), thus in total the same number of weights as the base architecture. Adding longitude and latitude information, surprisingly, deteriorates forecast performance for z500, while it slightly increases forecast skill of t850 at intermediate lead times (second-to-last bars in Fig. 3). We however note that we have not tuned our model for these additional inputs, and we cannot exclude that the number of parameters and therefore the capacity of the network is not large enough to exploit this additional information. The results with the hemispheric convolution with same number of weights as the base architecture are also intriguing (last bars in Fig. 3). The forecast skill deteriorates compared to the base architecture, except for ACC at longer lead times, while for RMSE it detoriates at all lead times. This result shows that the skill increase brought by the non-shared hemispheric convolution reported above actually seems to come from the higher number of parameters, and not from the separate convolutions. Conversely, the improved skill seen in the architecture with shared but flipped parameters clearly shows that the flipping provides an advantage.

\begin{table}
\caption{\label{tab:baseline-scores}RMSE at days 3 and 5, for all architectures, including NWP model scores as baselines. High-resolution is abbreviated as ``hres'', low-resolution as ``lres''. The numbers in parentheses denote the standard deviation of the errors between the 4 training realizations of each network. The lowest RMSE for
each variable, lead time and resolution (not considering IFS) is highlighted in bold.}
\begin{tabular}{c|p{2cm}|p{2cm}|p{2cm}|p{2cm}}
\hline 
 & geopotential 500hPa day 3  $[m2/s2]$ & geopotential 500hPa day 5 $[m2/s2]$ & temperature 850hPa day 3 $[K]$ & temperature 850hPa day 5 $[K]$\\
\hline 
\hline 
 hres base & 542 (7) & 845 (40) & 2.87 (0.09) & 4.11 (0.16)\\
\hline 
 hres hemconv & 519 (15) & 785 (24) & 2.73 (0.07) & 3.83 (0.11)\\
\hline 
 hres hemconv\_sharedweights & 492 (14) & 767 (26) & 2.61 (0.02) & 3.68 (0.10)\\
\hline 
 hres sphereconv & 487 (6) & 773 (8) & 2.64 (0.07) & 3.73 (0.13)\\
\hline 
 hres sphereconv\_hemconv & 455 (13) & 736 (28) & 2.57 (0.02) & 3.68 (0.04)\\
\hline 
 hres sphereconv\_hemconv\_shared & \textbf{415 (6)} & \textbf{696 (11)} & \textbf{2.45 (0.05)} & \textbf{3.53 (0.08)}\\
\hline 
 lres base & 514 (14) & 788 (26) & 2.75 (0.08) & 3.81 (0.12)\\
\hline 
 lres base\_latlon & 520 (17) & 806 (34) & 2.73 (0.03) & 3.82 (0.07)\\
\hline 
 lres hemconv & 490 (4) & 773 (6) & 2.69 (0.07) & 3.80 (0.10)\\
\hline 
 lres hemconv\_halfsize & 535 (12) & 819 (27) & 2.82 (0.11) & 3.88 (0.16)\\
\hline 
 lres hemconv\_sharedweights & 458 (18) & 737 (33) & 2.55 (0.05) & 3.63 (0.07)\\
\hline 
 lres sphereconv & 480 (13) & 762 (20) & 2.62 (0.08) & 3.70 (0.14)\\
\hline 
 lres sphereconv\_hemconv & 455 (6) & 727 (12) & 2.56 (0.02) & 3.62 (0.02)\\
\hline 
 lres sphereconv\_hemconv\_shared & \textbf{416 (14)} & \textbf{699 (35)} & \textbf{2.50 (0.07)} & \textbf{3.61 (0.09)}\\
\hline 
IFS T42 & 489 & 743 & 3.09 & 3.83\\
\hline 
IFS T63 & 268 & 463 & 1.85 & 2.52\\
\hline 
Operational IFS & 154 & 334 & 1.36 & 2.03\\
\hline 
\end{tabular}
\end{table}

\subsection{Analysis of events with largest forecast errors}

\begin{figure}
\includegraphics[width=1\textwidth]{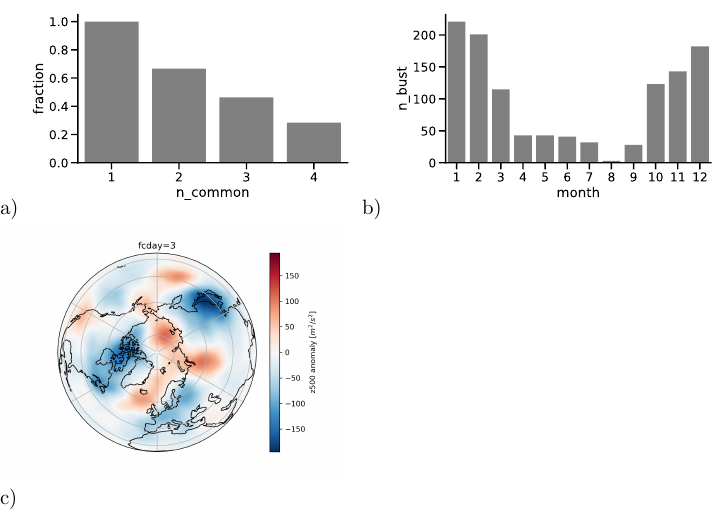}

\caption{\label{fig:extremes}a) Fraction of dates with extreme forecast error
(RMSE $>$95\%) in the Northern Hemisphere (NH) that occur in at least n\_common out of the 4 training realisation for sphereconv\_hemconv\_shared.
b) Monthly count of days with at least one training realisation with extreme forecast
error in the NH. c) Composite anomalies of z500
on all initialisation dates for which at least one training realisation has an extreme forecast error at leadtime 3 days in the NH. The anomalies are computed relative to deseasonalized (monthly) means. All data is from the high resolution forecasts over the period 2017--2018.}

\end{figure}

We now look at the forecasts within the upper 5\% of RMSE (forecast ``busts'') for the Northern Hemisphere (NH) for lead-time 3 days for
the spherconv\_hemconv\_shared architecture (the architecture which, as discussed in Sect. 3\ref{sect:evaluation}, generally displays the highest forecast skill). For each
of the 4 training realizations, the percentile is computed individually.
When comparing the initialization dates of the worst forecasts, \textasciitilde 28\%
are exactly the same dates for all training realisations, \textasciitilde 46\%
occur in at least 3 of the 4 members, and \textasciitilde 66\% in at
least 2 of the 4 members (Fig. \ref{fig:extremes} a). This is much
higher than expected by chance if the events were randomly distributed.
Events with large errors are more common in boreal winter than in summer (Fig.
\ref{fig:extremes} b), in line with the performance of operational
NWP models. Similarly, repeating the analysis on Southern Hemisphere data shows that events with large errors are more common in austral winter (Fig. A2).

Finally, Fig. \ref{fig:extremes}c) shows composite z500 anomalies
at all initialisation times for which at least one of the spherconv\_hemconv\_shared
high-resolution networks had an error $>$ 95\% in the NH. The anomaly is computed
with respect to the mean over 2017--2018 (the evaluation period), and
separately for each month. There are positive anomalies east of Greenland, in central Russia
and in the middle of the North Pacific, and negative
anomalies in northern Canada, the eastern coast of Asia and
central Europe. There is also an anomaly dipole bewteen the west coast of North America and the eastern Pacific, at lower latitudes than the other anomaly centres. These collectively constitute a circumhemispheric wave-4 pattern. The results are qualitatively similar for other architectures
(Figs. A3--A5), and for other lead-times (Figs. A6--A9). This indicates
that the skill of the network forecasts is dependent on the atmospheric
configuration, just as in NWP forecasts (e.g. \citet{ferranti_flow-dependent_2015,matsueda2018estimates}).

\subsection{Computational Performance}

Replacing standard convolution with Spherenet convolution introduces
a significant amount of additional computations. While the use of
sparse tensorflow tensors (see Sect. 2.b) for the interpolation tensor $\hat{L}$
avoids large memory requirements, the computation time of the sphereconv
network for the high-resolution data compared to the base architecture is roughly
a factor 8 higher on a CPU with 2 cores (9.3 vs 1.1s), and by a factor
of 150 higher on a NVIDIA Tesla v100 GPU (3.3s vs 23ms). The computational overhead might be reduced by optimizing the implementation for certain computational architectures, such as in \citet{esteves2023scaling}, who optimized their code for Tensor Processing Units (TPUs). 
Using hemisphere-wise
convolution, on the other hand, does not introduce any significant performance overhead
(with shared weights $\sim$4\%, with separate weights none
at all).

\section{Discussion and Conclusion}

In this paper, we have tested two approaches to improve data-driven
weather forecasts with CNNs. The aim was not to develop the best possible neural network
based weather forecasts (and indeed we note that there are publications presenting data-driven forecasts with a higher skill than ours), but to assess the effect of specific
changes to a conventional neural network architecture on forecast skill. Firstly, we have tested replacing standard
convolution operations with a convolution operation that takes into account the (near-)spherical shape of the Earth. Secondly,
we tested integrating basic meteorological knowledge into the structure
of the networks, namely that the dynamics of the two hemispheres are different. This is done in two ways: in the first case, we hardcode into the network that the dynamics of one hemisphere are ``flipped'' with respect to the other  hemisphere. This is implemented by flipping the weights of the network. In the second case, we use independent weights for each hemisphere, thus leaving the network free to learn potential differences in dynamics.
These methods (and combinations of them) were tested on the ERA5 data
from the WeatherBench dataset \citep{rasp_weatherbench_2020}. We
used a neural network architecture previously proposed by \citet{weyn_improving_2020},
and adapted it to our convolution methods. We found that both the
Spherenet convolution and the hemispheric information improve the forecasts,
but in subtly different ways. Both Spherenet convolution and flipped hemispheric information lead to the largest improvements
close to the poles, and less so in other regions. However, adding the Spherenet convolution to the flipped hemisphere-specific
information leads to relatively uniform improvements in the
mid-latitudes when compared to hemispheric information alone. The mid-latitudes are the regions where the largest forecast errors appear in the first place. For most lead times, skill metrics, resolutions and variables considered here, combining Spherenet convolution with hemisphere-specific information using shared flipped weights leads to the best forecasts. 
The experiments with hemispheric convolution without weight sharing show some interesting results. While hemispheric convolution without weight sharing showed a skill increase, the additional experiment with keeping the number of parameters the same as in the base architecture showed that this skill increase actually seems to come mainly from the additional paramaters, and not the hemispheric convolution.
This however clearly shows that the skill increase in the hemispheric convolution \emph{with} shared convolution comes from the weight-sharing and flipping, showing the value of this physics-inspired approach.

Finally, we have found that initial conditions causing the largest forecast
errors are relatively consistent across different training realizations
of the same network, and across different network architectures and lead times. This indicates that, as for conventional NWP forecasts, forecast errors of the neural networks are at least partly flow-dependent. In other words, the networks struggle to make skilful forecasts when initialised from specific atmospheric states. Previous work has shown that quantifications of an ``intrinsic'' atmospheric predictability only partly match the empirical forecast errors from a NWP initialised from that atmospheric state \citep{scher_predicting_2018-1,hochman2019new,hochman2021new,hochman2022dynamics}. We do not investigate here whether the atmospheric configurations leading to high forecast errors for the CNNs used here match configurations with low `intrinsic'' atmospheric predictability. Indeed, the error could also depend on the error in the reanalysis product used for training or on deficiencies in the training of the networks themselves.

Possible improvements
to the methods presented here could be:
\begin{itemize}
\item To split up the convolution for smaller regions instead of at hemispheric scale (similar
to \citet{weyn_improving_2020}), which could lead to the shared flipped weights no longer presenting a forecast skill advantage over independent convolutions.
\item To include locally connected layers. Here each gridpoint in a layer
is also a combination of the inputs from a certain kernel (e.g. 3$\times$3),
but the weights are not shared across the domain.
\item To add more prior information into the structure of the network, for
example on the vertical structure of the atmosphere.
\end{itemize}

In this study, we used an existing architecture, and changed the convolution types. Due to limited computational resources, we did not perform any additional hyperparameter tuning. An additional methodological improvement would therefore be to tune each individual architecture separately. Related to this, one could also compare scores for the training set both between architectures and to those on the test set, to verify whether any over or under fitting in occurring for the different models.

Our approach differs from previous studies in the field in several
respects. The closest previous work is that by \citet{weyn_improving_2020}, who split up the world into
six regions --- two polar and four equatorial regions --- with each region being represented by a local
grid. On these local grids they used standard convolution operations.
This method still leads to distortions, as even a subregion of the
Earth's surface cannot be represented on a local regular grid with
complete accuracy. In addition, this method also needs padding at the edge of each of the 6 regions, which introduces some ambiguity at the corners. \citet{weyn_improving_2020} further use two sets of weights, one for all four equatorial regions and the other one for the two polar regions. The weights for the polar regions are ``flipped'' between the two poles. This is similar to our idea of flipped weights to provide hemispheric information to the CNN. 
There are also several differences between the approach we present here and that of \citet{weyn_improving_2020} from a practical point of view.
The method of \citet{weyn_improving_2020} needs data-preprocessing (regridding), but can then use standard neural network operations. In our approach,
the standard data can be used, but the Spherenet convolution introduces a
computational overhead in every pass through the network.

The percent increase in runtime needed for the networks
with Spherenet convolution is higher on a GPU than on CPU, which could indicate
that the implementation using sparse tensorflow tensors -- which we adopt here -- is not optimized
for GPUs. For small input sizes, an alternative would be to use standard
tensorflow tensors (still filled sparsely, but represented as a full
tensor (array) in memory). However, for the full resolution ERA5 data (0.25$^\circ$
resolution, namely 3600x1801 gridpoints on a regular lat-lon points), this
would not be feasible with current computers due to memory limitations. Indeed, the interpolation tensor would then have a size of 6483600x6483600.

The methods used and presented in this study all generate single deterministic forecasts. In many weather forecasting settings, however, probabilistic forecasts are wanted. The typical way to do this with numerical weather prediction models is to use ensemble forecasting, which entails performing multiple model runs with slightly
different initial conditions, slightly different model formulations, stochastic components,
or a combination of these (e.g. \citep{leutbecher2008ensemble}). While dedicated probabilistic ML techniques exist, such as gaussian process regression \citep{ebden2015gaussian} or probabilistic neural networks \citep{MOHEBALI2020347}, the at least conceptually simpler approach is to apply the concept of ensemble NWP to ML models as well. \cite{scher_ensemble_2020} compared three methods of generating ensembles from deterministic neural network prediction systems. The first one -- using multiple models, each trained with different initial seeds, has a computational cost which scales linearly with ensemble size both in the training stage and the prediction stages. Using random initial perturbations requires no additional training, and scales linearly in the prediction phase with no significant additional overhead, as generating the random perturbations is computationally cheap. The final and most advanced method uses singular value decomposition to find optimal initial perturbations. This introduces a one-time overhead per prediction for generating the singular vectors for the initial perturbations, and apart from this the computational cost again scales linearly with ensemble size. In principle, the improvements to conventional CNNs proposed here would be combined with any of the above approaches, although the one requiring to train the model multiple times may become computationally demanding given the additional overhead from the Spherenet convolution.

To conclude, in this study we have tested some simple improvements to conventional convolutional neural network architectures for weather forecasting. These are based on a
convolution operator accounting for the Earth's (near-)spherical geometry and on providing the network with knowledge that the climate dynamics of the two hemispheres differ. 
We did not seek to outperform state-of-the-art data-driven weather forecasts in terms of forecast skill but instead sought to test the limits of CNNs with a relatively small number of parameters. This can be beneficial for research groups who may not have enough computational capacity to run very large transformer or GNN models. The improvements presented here stem from physical reasoning, and may support future development of physics-oriented ML weather prediction models.

\clearpage
%%%%%%%%%%%%%%%%%%%%%%%%%%%%%%%%%%%%%%%%%%%%%%%%%%%%%%%%%%%%%%%%%%%%%
% ACKNOWLEDGMENTS
%%%%%%%%%%%%%%%%%%%%%%%%%%%%%%%%%%%%%%%%%%%%%%%%%%%%%%%%%%%%%%%%%%%%%
\acknowledgments
S.S. was funded by the Dept. of Meteorology of Stockholm University. G.M. was supported by the European Union’s H2020 research and innovation programme under ERC grant no. 948309 (CEN\AE project) and by the Swedish Research Council Vetenskapsr\aa det (grant no.: 2016--03724).
This work was partly supported by the ``DDAI'' COMET Module within
the COMET -- Competence Centers for Excellent Technologies Programme, funded by the Austrian Federal Ministry (BMK and BMDW), the Austrian Research Promotion Agency (FFG), the province of Styria (SFG) and partners from industry and academia. The COMET Programme is managed by FFG.
The computations and data handling were enabled by resources provided by the National Academic Infrastructure for Supercomputing in Sweden (NAISS) and the Swedish National Infrastructure for Computing (SNIC) at the High Performance Computing Center North (HPC2N) and National Supercomputer Centre (NSC), partially funded by the Swedish Research Council through grant agreements no. 2022-06725 and no. 2018-05973.
%  Keep acknowledgments (note correct spelling: no ``e'' between the ``g'' and
% ``m'') as brief as possible. In general, acknowledge only direct help in
%  writing or research. Financial support (e.g., grant numbers) for the work done, 
%  for an author, or for the laboratory where the work was performed must be 
%  acknowledged here rather than as footnotes to the title or to an author's name.
%  Contribution numbers (if the work has been published by the author's institution 
%  or organization) should be placed in the acknowledgments rather than as 
%  footnotes to the title or to an author's name.

%%%%%%%%%%%%%%%%%%%%%%%%%%%%%%%%%%%%%%%%%%%%%%%%%%%%%%%%%%%%%%%%%%%%%
% DATA AVAILABILITY STATEMENT
%%%%%%%%%%%%%%%%%%%%%%%%%%%%%%%%%%%%%%%%%%%%%%%%%%%%%%%%%%%%%%%%%%%%%
% 
%
\datastatement
The software developed for this study and the intermediate data underlying the plots are available on the first Authors github repository \url{https://github.com/sipposip/physics-informed-ML-NWP} as well as on zenodo \url{https://zenodo.org/record/8344872} with doi \url{10.5281/zenodo.8344872}. Additionally, the trained models are available stored in the zenodo archive as well.
The ERA5 WeatherBench data used as input data can be freely obtained via WeatherBench
\url{https://github.com/pangeo-data/WeatherBench}
and alternatively as raw data freely from the Copernicus data store
at \url{https://cds.climate.copernicus.eu/cdsapp\#!/dataset/reanalysis-era5-single-levels?tab=overview}
%  The data availability statement is where authors should describe how the data underlying 
%  the findings within the article can be accessed and reused. Authors should attempt to 
%  provide unrestricted access to all data and materials underlying reported findings. 
%  If data access is restricted, authors must mention this in the statement. See
%  {http://www.ametsoc.org/PubsDataPolicy} for more info.

%%%%%%%%%%%%%%%%%%%%%%%%%%%%%%%%%%%%%%%%%%%%%%%%%%%%%%%%%%%%%%%%%%%%%
% APPENDIXES
%%%%%%%%%%%%%%%%%%%%%%%%%%%%%%%%%%%%%%%%%%%%%%%%%%%%%%%%%%%%%%%%%%%%%
%
%% If only one appendix, use

\appendix

%% If more than one appendix, use \appendix[<letter>], e.g.,

%\appendix[A] 

%% Appendix title is necessary! For appendix title:

\appendixtitle{Additional analyses}

\begin{figure}
\includegraphics[width=0.9\textwidth]{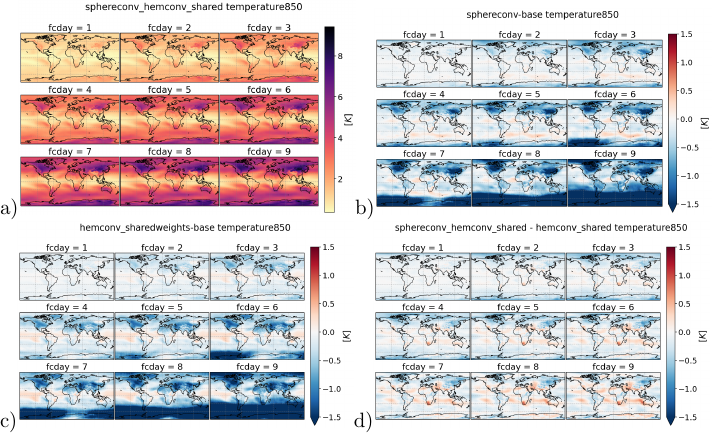}
\caption{As Fig. 4, but for t850 {[}K{]}}
\end{figure}

\begin{figure}
\includegraphics[width=0.9\textwidth]{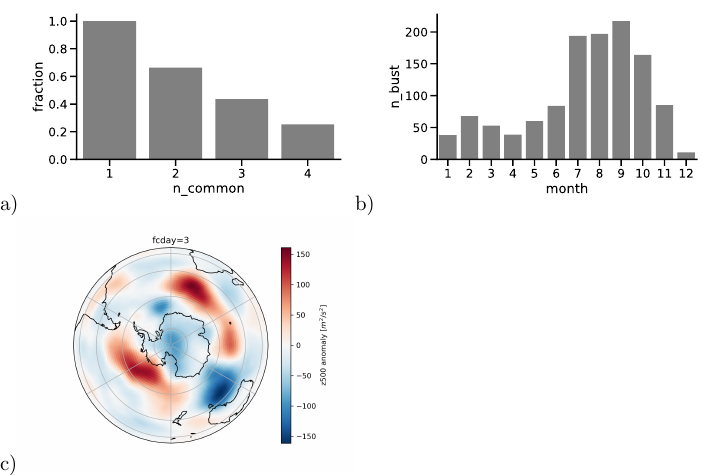}
\caption{\label{fig:extremes-2-1-1-1}As Fig. 5, but for the Southern Hemisphere}
\end{figure}

\begin{figure}
\includegraphics[width=0.9\textwidth]{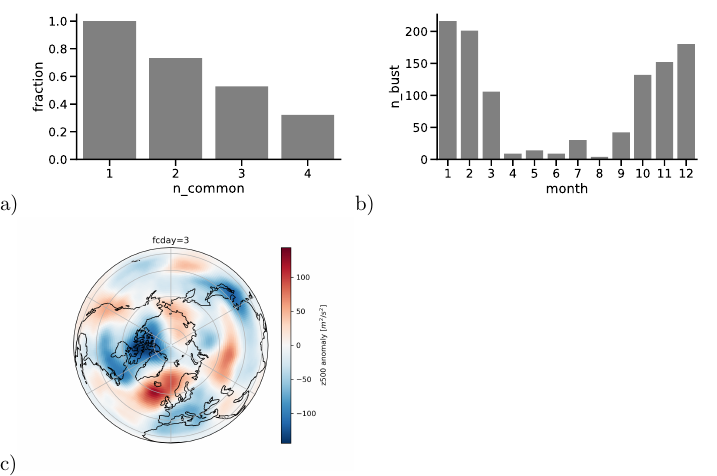}
\caption{\label{fig:extremes-1}As Fig. 5, but for the base architecture.}
\end{figure}

\begin{figure}
\includegraphics[width=0.9\textwidth]{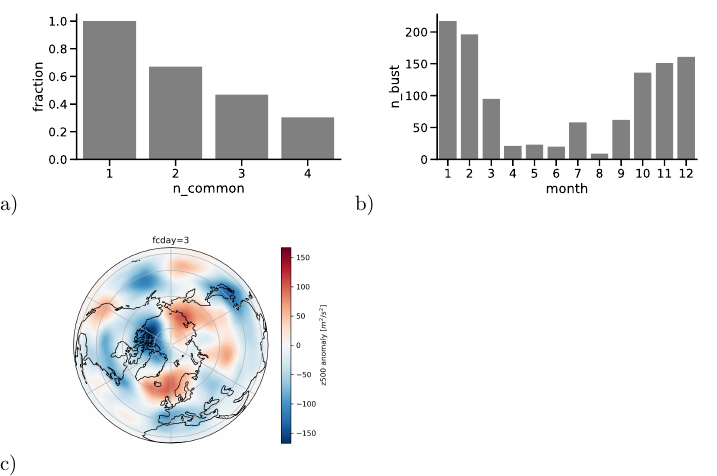}
\caption{\label{fig:extremes-1-1}As Fig. 5, but for the hemconv\_shared
architecture.}
\end{figure}

\begin{figure}
\includegraphics[width=0.9\textwidth]{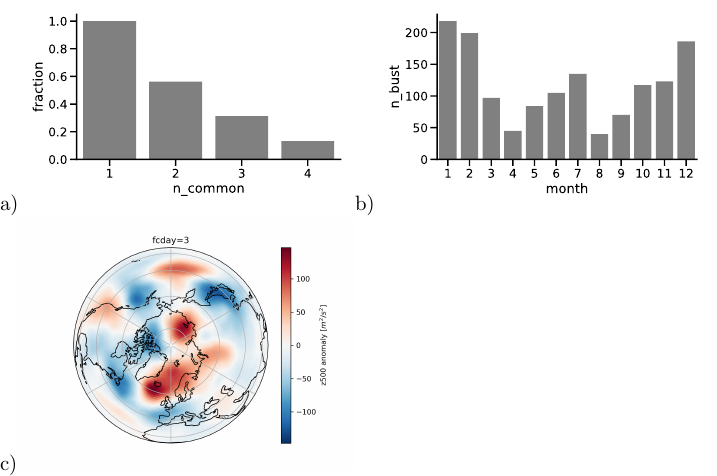}
\caption{\label{fig:extremes-1-2}As Fig. 5, but for the sphereconv architecture.}
\end{figure}

\begin{figure}
\includegraphics[width=0.9\textwidth]{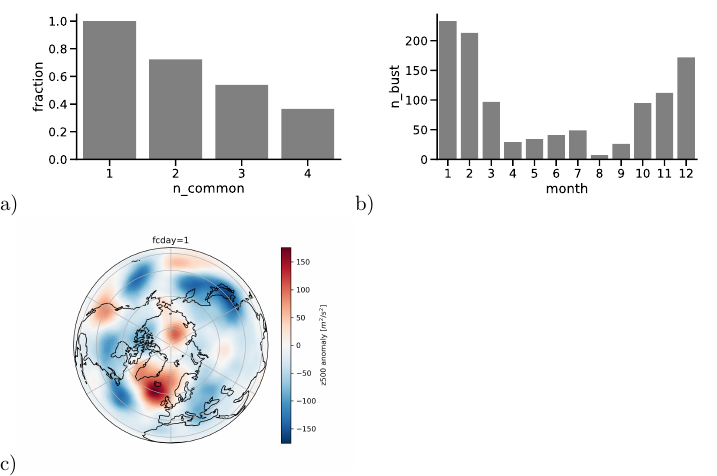}
\caption{\label{fig:extremes-2}As Fig. 5, but for lead time of 1 day.}
\end{figure}

\begin{figure}
\includegraphics[width=0.9\textwidth]{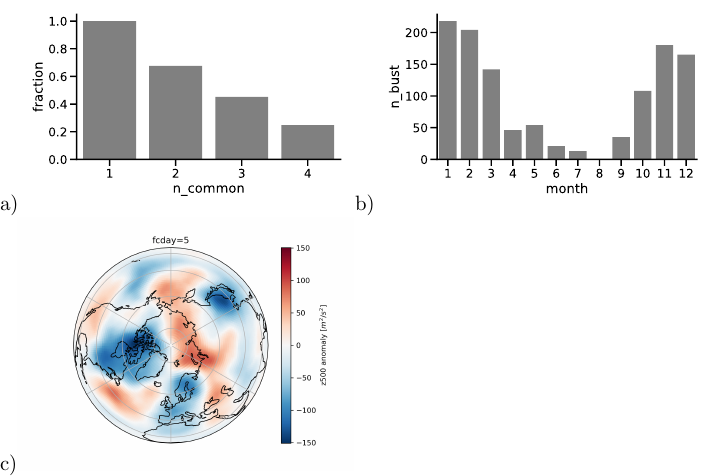}
\caption{\label{fig:extremes-2-1}As Fig. 5, but for lead time of 5 days.}
\end{figure}

\begin{figure}
\includegraphics[width=0.9\textwidth]{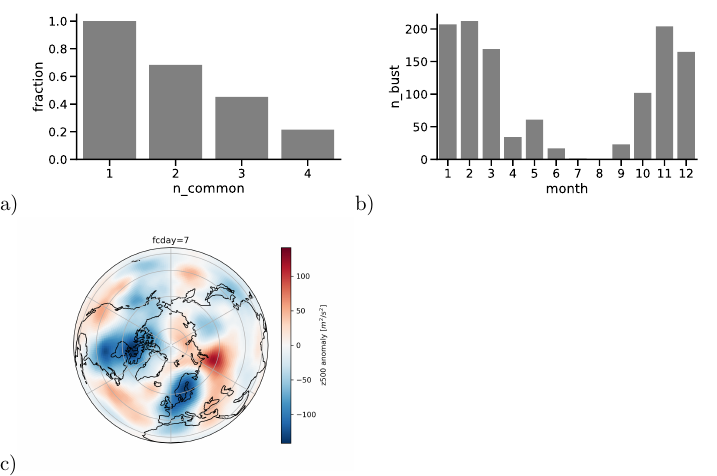}
\caption{\label{fig:extremes-2-1-1}As Fig. 5, but for lead time of 7 days.}
\end{figure}

\begin{figure}
\includegraphics[width=0.9\textwidth]{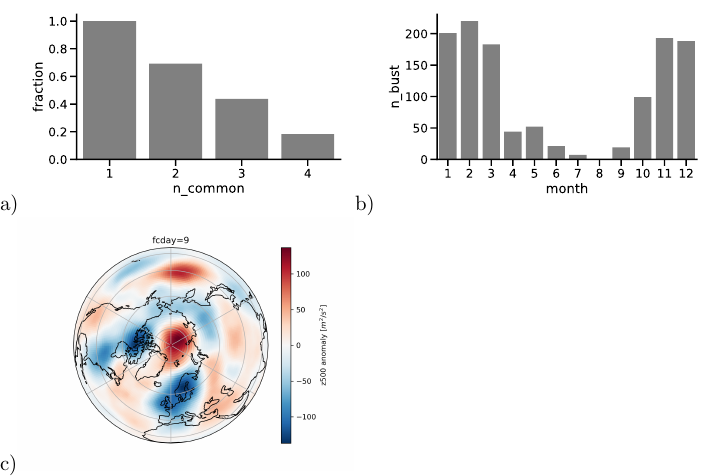}
\caption{\label{fig:extremes-2-1-1-1}As Fig. 5, but for lead time of 9 days.}
\end{figure}

%%% Appendix section numbering (note, skip \section and begin with \subsection)
%
% \subsection{First primary heading}

% \subsubsection{First secondary heading}

% \paragraph{First tertiary heading}

\clearpage
%%%%%%%%%%%%%%%%%%%%%%%%%%%%%%%%%%%%%%%%%%%%%%%%%%%%%%%%%%%%%%%%%%%%%
% REFERENCES
%%%%%%%%%%%%%%%%%%%%%%%%%%%%%%%%%%%%%%%%%%%%%%%%%%%%%%%%%%%%%%%%%%%%%
% Make your BibTeX bibliography by using these commands:
% \bibliographystyle{ametsocV6}
% \bibliography{references}
\bibliographystyle{ametsocV6}
\bibliography{nn_fc_spherconv}
\end{document}